\documentclass{ws-procs975x65}

\def\beq{\begin{equation}}
\def\eeq{\end{equation}}

\begin{document}

\title{Relativistic and Newtonian fluid tori with electric charge}
\author{Vladim\'{\i}r Karas and Ond\v{r}ej Kop\'a\v{c}ek}
\address{Astronomical Institute, Czech Academy of Sciences,\\
Bo\v{c}n\'{\i} II 1401, CZ-14100 Prague, Czech Republic\\
E-mail: vladimir.karas@cuni.cz}

\author{Audrey Trova} 
\address{Center of Applied Space Technology and Microgravity (ZARM), University of Bremen,\\ 
Am Fallturm, D-28359 Bremen, Germany}

\author{Ji\v{r}\'{\i} Kov\'a\v{r}, Petr Slan\'y and Zden\v{e}k Stuchl\'{\i}k}
\address{Institute of Physics, Faculty of Philosophy and Science, Silesian University in Opava,\\ 
Bezru\v{c}ovo n\'am.\ 13, CZ-74601 Opava, Czech Republic}

\begin{abstract}
We discuss the effects of electric charging on the equilibrium configurations of magnetized, rotating fluid tori around black holes of different mass. In the context of gaseous/dusty tori in galactic nuclei, the central black hole dominates the gravitational field and it remains electrically neutral, while the surrounding material acquires some electric charge and exhibits non-negligible self-gravitational effect on the torus structure. The structure of the torus is influenced by the balance between the gravitational and electromagnetic forces. A cusp may develop even in Newtonian tori due to the charge distribution.
\end{abstract}
\bodymatter

\section{Introduction}

Within the framework of General Relativity, black holes are fully described by a small number of parameters
\cite{Car:1968}. The most relevant from the viewpoint of astrophysical applications are the mass of the black 
hole, $M_\bullet$, and angular momentum $J_\bullet$ or, in a dimension-less form, the spin parameter
$a^*\equiv a/M_\bullet=J_\bullet c/GM_\bullet^2$. The classical black-hole solution of Kerr-Newmann can be equipped
by electric charge, although any significant value appears to be astrophysically unlikely 
over an extended duration because of continued process of neutralization by a selective charge accretion 
from the surrounding plasma into any cosmic black holes must be embedded.

The Kerr-Newman solution the line element adopts the form
adopts the form \cite{Mis-Tho-Whe:1973}
\begin{equation}
\label{metKN}
ds^2=-\frac{\Delta}{\Sigma}\left(dt-a\sin{\theta}d\phi\right)^2+\frac{\Sigma}{\Delta}dr^2+\Sigma d\theta^2+\frac{\sin^2{\theta}}{\Sigma}\left[\left(r^2+a^2\right)d\phi-adt\right]^2,
\end{equation}  
in dimensionless Boyer-Linquist coordinates $(t,r,\phi,\theta)$, where $\Delta(r)=r^2-2r+a^2+e^2$ and $\Sigma(r,\theta)=r^2+a^2\sin^2{\theta}$. Quantities $a$ and $e$ are the rotational and charge parameters of the spacetimes. The associated non-zero components of the antisymmetric electromagnetic field tensor $F_{ij}=A_{j,i}-A_{i,j}$ are
\begin{eqnarray}
\label{elamg1}
F_{rt}&=&\frac{e(r^2-a^2\cos^2{\theta})}{\Sigma^2},\qquad
\label{elmag2}
F_{r\phi}=\frac{-ae\sin^2{\theta}(r^2-a^2\cos^2{\theta})}{\Sigma^2},\\
\label{elmag3}
F_{\theta t}&=&\frac{-a^2er\sin{2\theta}}{\Sigma^2},\qquad
\label{elmag4}
F_{\theta\phi}=\frac{aer\sin{2\theta}(r^2+a^2)}{\Sigma^2}.
\end{eqnarray}

The effective potential $W(r,\theta)$ for the motion of a particle with the specific charge $\tilde{q}=q/m$ satisfies the relation 
\begin{eqnarray}
\label{potential}
XW=Y+\sqrt{Y^2-XZ},
\end{eqnarray}
where the functions
$X(r,\theta)=(r^2+a^2)^2-\Delta a^2\sin^2{\theta}$,
$Y(r,\theta)=(\tilde{L}a+\tilde{q}er)(r^2+a^2)-\tilde{L}a\Delta$, and
$Z(r,\theta)=(\tilde{L}a+\tilde{q}er)^2-\Delta\Sigma-\Delta\tilde{L}^2/\sin^2{\theta}$;
here, $\tilde{L}=L/m$ is the conserved axial component of specific angular momentum. The 
effective potential (\ref{potential}) develops local minima outside the equatorial plane $\theta=\pi/2$,
which suggests the possibility of halo orbits.\footnote{Within the particle approximation the halo orbits are formed by stable bound 
trajectories which do not cross the equator. In the fluid analogy they correspond to toroidal structures
with the pressure maxima at $\theta\neq\pi/2$ (i.e.\ $z\neq0$).\cite{dul1,kov1}}
The minima occur provided that the electric charge of the accreted matter is non-zero and interacts with
the global magnetic field of the central object and/or of the external origin due to currents flowing
outside the horizon.

\begin{figure}[tbh!]
\begin{center}
\vspace*{0em}
\includegraphics[width=0.8\textwidth]{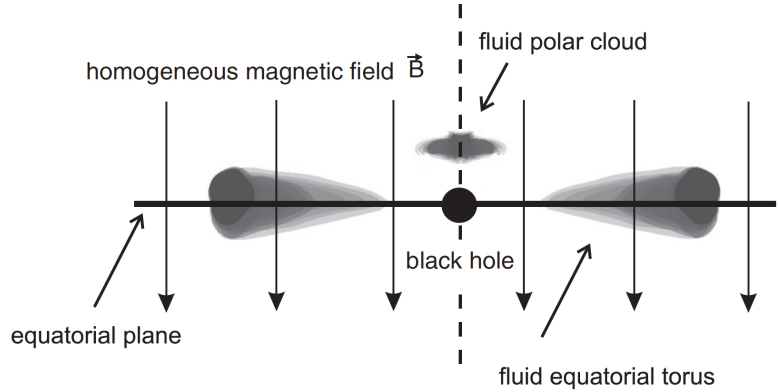}
\end{center}
\vspace*{0em}
\caption{A black hole is immersed in asymptotically homogeneous magnetic field $\vec{B}$
and encircled by a torus-like equatorial fluid configuration, and an off-equatorial polar cloud.
The charge distribution within the fluid interacts with the magnetic field and helps to levitate the fluid
above the equatorial plane. Similar effect occurs for trajectories of electrically charged particles.}
\label{fig1}
\end{figure}

In the other words, the stable configuration allows for a distribution of charged matter near the equator, 
in lobes located symmetrically above and below the equatorial plane, or even around the polar axis (see Fig. \ref{fig1}).
This represents general-relativistic version of St\"{o}rmer's halo orbits that have been 
explored in connection with the motion of electrically charged dust grains in planetary magnetospheres.\cite{sto1}
In the case of Kerr-Newman solution, it can be found that the particular form of mutually connected gravitational 
and electromagnetic fields do not allow existence of stable halo orbits above the outer horizon;
an additional external magnetic field is required.

The prevailing process of neutralization, however, may not prevent the surrounding material to 
undergo electric charging in the environment of complex plasma that is irradiated
by energetic X-rays, as is often the case in active galactic nuclei. This mechanism leads to charge separation 
within the dusty tori. Therefore, we have proposed the process of charging as mechanism that can
maintain geometrical thickness of the dusty tori in the vertical direction.\cite{kov2}

Furthermore,
the existence of mutually detached regions (lobes) of bound stable orbits can provide a framework 
to trigger oscillations; it allows us to imagine a situation when the bulk motion along the 
circular trajectories is superposed by a fraction o material that performs oscillatory
motion between the disconnected lobes. Halo orbits could have outstanding consequences for
the radiation signal from accreting black holes. On the other hand, 
general relativity effects have a tendency to bring the halo orbits closer to the equatorial plane compared
to corresponding Newtonian solutions. Also,
to ensure the intrinsic self-consistency of the model we need to include self-gravity of the
medium in the consideration, as this can further diminish the vertical thickness
of the figures of equilibrium of electrically charged fluid.\cite{kar1,aud1,aud2,aud3}

\section{Electrically charged tori near a magnetized black hole}
We describe the fluid by polytropic equation of state, $p=\kappa \rho^{\Gamma}$,
with $\kappa$ and $\Gamma$ being parameters of the polytrope. This relation ensures conservation of entropy, as is appropriate for a perfect fluid. Neglecting the electrostatic corrections to the equation of state we can use the polytropic equation consistently even to described an electrically charged state. In the case of sufficiently small $\rho$ the medium is non-relativistic and the contribution of the specific internal energy $\epsilon/\rho-1$ to the total energy density becomes negligible (i.e., $\epsilon \approx \rho$), however, this assumption is not accurate enough once the density increases; in such a case the self-electromagnetic field cannot be ignored.

The configuration of the fluid is determined by its rotation in the \mbox{$\phi$-direction} with \mbox{4-velocity} $U^{\alpha}=(U^t,U^{\phi},0,0)$, specific angular momentum $\ell=-U_{\phi}/U_t$ and angular velocity $\omega=U^{\phi}/U^t$. These are related by
\begin{equation}
\label{Omega}
\omega=-\frac{\ell g_{tt}+g_{t\phi}}{\ell g_{t\phi}+g_{\phi\phi}},\quad
(U_t)^2=\frac{g_{t\phi}^2-g_{tt}g_{\phi\phi}}{\ell^2 g_{tt}+2\ell g_{t\phi}+g_{\phi\phi}}.
\end{equation}
The shape of a charged fluid torus with a charge density profile $q$ and energy density $\epsilon$ is determined by isobaric surfaces of the pressure $p$ profile (equi-pressure surfaces), which can be derived from the conservation equation
 \begin{eqnarray}
\label{master}
\nabla_{\beta}T^{\alpha\beta}=F^{\alpha\beta}J_{\beta},
\end{eqnarray}
where $J^{\alpha}$ is the fluid \mbox{4-current} density and $T^{\alpha\beta}$ is the stress-energy tensor,  
\begin{eqnarray}
\label{Tmat}
T^{\alpha\beta}&=&(\epsilon+p)U^{\alpha}U^{\beta}+pg^{\alpha\beta}.
\end{eqnarray}
From the above-given relation two partial differential equations follow for the radial and latitudinal distribution of pressure,
\begin{equation}
\label{pressure}
\frac{\partial p}{\partial r}=-(p+\epsilon)\mathcal{R}_1+q\mathcal{R}_2\equiv \mathcal{R}_0,\quad
\frac{\partial p}{\partial \theta}=-(p+\epsilon)\mathcal{T}_1+q\mathcal{T}_2\equiv \mathcal{T}_0,
\end{equation}
where $\mathcal{R}_0=\mathcal{R}_0(r,\theta)$ and $\mathcal{T}_0=\mathcal{T}_0(r,\theta)$ denote the right hand sides of the two latter 
equations.\cite{kov2}

As an specific example of the imposed electromagnetic field we can write the static case of an asymptotically uniform magnetic field with intensity $B$. Such settings can be described by the special case (zero rotational parameter) of Wald's test-field solution of Einstein-Maxwell equations, which in dimensionless units reads 
\begin{eqnarray}
\label{wald}
A_t=-e/r,\quad
A_{\phi}=\textstyle{\frac{1}{2}}\,Br^2\sin^2{\theta},
\end{eqnarray}
describing the electromagnetic field in the background of Schwarzschild geometry. A generalization of the adopted approach to the case of a rotating (Kerr) black hole is straightforward but technically more complicated and beyond the limited space of the present contribution.

\section{Conclusions}
We further explored off-equatorial configurations of electrically charged fluid near magnetized black holes.
By comparison with previous works we showed that the conditions of existence of these configurations can be diverse. 
By the mass-scaling paradigm for cosmic black-hole sources,
similar effects can occur also in the context of stellar-mass black holes, where a massive torus can form as a remnant 
of a tidally disrupted companion.

\section*{Acknowledgments}
V.K. and A.T. acknowledge the Czech Science Foundation (GA\v{C}R) -- 
German Research Foundation (DFG) collaboration project No.\ 19-01137J.

\end{document}